# CONTINUUM GATING CURRENT MODELS COMPUTED WITH CONSISTENT INTERACTIONS

Tzyy-Leng Horng[1], Robert S. Eisenberg[2], Chun Liu[3], Francisco Bezanilla[4]

[1]Department of Applied Mathematics, Feng Chia University, Taichung, Taiwan 40724
[2]Department of Molecular Biophysics and Physiology, Rush University, Chicago, Illinois 60612, United States
[3]Department of Mathematics, Pennsylvania State University, University Park, Pennsylvania 16802, United States
[4]Department of Biochemistry and Molecular Biology, University of Chicago, Chicago, Illinois 60637, United States

**Abstract**

The action potential signal of nerve and muscle is produced by voltage sensitive channels that include a specialized device to sense voltage. Not surprisingly the voltage sensor depends on the movement of permanent charges of basic side chains in the changing electric field, as suggested long ago by Hodgkin and Huxley and Bezanilla and Armstrong, for example. Gating currents of the voltage sensor are now known to depend on the back-and-forth movements of positively charged arginines through the hydrophobic plug of a voltage sensor domain. Transient movements of these permanently charged arginines, caused by the change of transmembrane potential, further drag the S4 segment and induce opening/closing of ion conduction pore by moving the S4-S5 linker. The ion conduction pore is a separate device from the voltage sensor, linked (in an unknown way) by the mechanical motion and electric field changes of the S4-S5 linker. This moving permanent charge induces capacitive current flow everywhere. Everything interacts with everything else in the voltage sensor so everything must interact with everything else in its mathematical model, as everything does in the whole protein. A PNP-steric model of arginines and a mechanical model for the S4 segment are combined using energy variational methods in which all movements of charge and mass satisfy conservation laws of current and mass. The resulting 1D continuum model is used to compute gating currents under a wide range of conditions, corresponding to experimental situations. Conservation laws are partial differential equations in space and time. Chemical-reaction-type models based on ordinary differential equations cannot capture such interactions with one set of parameters. Indeed, they may inadvertently violate conservation of current. Conservation of current is particularly important since small violations (<0.01%) quickly (<< $10^{-6}$ seconds) produce forces that destroy molecules. Our model reproduces signature properties of gating current: (1) equality of on and off charge in gating current (2) saturating voltage dependence in QV curve and (3) many (but not all) details of the shape of gating current as a function of voltage. The model computes gating current flowing in the baths produced by arginines moving in the voltage sensor. The model also captures the capacitive pile up of ions from the bulk solution in the form of electric double layers in vestibules adjacent to both ends of the hydrophobic plug. This pile up is coupled to the movement of arginines, and appears as a spike in the beginning of gating current that have been already observed as the early transient gating current in experiments. Our results agree qualitatively well with experiments, and can obviously be improved by adding more details of the structure and its correlated movements. The proposed continuum model is a promising tool to explore the dynamics and mechanism of the voltage sensor that avoids the sampling constraints that so limit the representation of biological concentrations ($10^{-7}$- $5\times10^{-1}$ M) and currents (time scale > $10^{-4}$ sec) in molecular dynamics simulations that compute atomic motion on the $10^{-15}$ sec time scale.

**1. Introduction**

Much of biology depends on the voltage across cell membranes. The voltage across the membrane must be sensed before it can be used by proteins. Permanent charges[1] move in the strong electric fields within membranes, so carriers of gating and sensing charge were proposed as voltage sensors even before membrane proteins were known to span lipid membranes [18-19, 21-23]. Movement of permanent charges of the voltage sensor is gating current and movement is the voltage sensing mechanism. Measurements of gating current were made much easier by biological preparations with artificially increased densities of voltage sensors and blocked conduction pores.

Knowledge of membrane protein structure has allowed us to identify and look at the atoms that make up the voltage sensor. Protein structures do not include the membrane potentials and macroscopic concentrations that power gating currents, and therefore simulations are needed. Atomic level simulations like molecular dynamics do not provide an easy extension from the atomic time scale $2\times10^{-16}$ sec to the biological time scale of gating currents that starts at $50\times10^{-6}$ sec and reaches $50\times10^{-3}$ sec. Calculations of gating currents from simulations must average the trajectories (lasting $50\times10^{-3}$ sec sampled every $2\times10^{-16}$ sec) of $\sim10^6$ atoms all of which interact through the electric field to conserve charge and current, while conserving mass. It is difficult to enforce continuity of current flow in simulations of atomic dynamics because simulations compute only local behavior while continuity of current is global, involving current flow far from the atoms that control the local behavior. It is impossible to enforce continuity of current flow in calculations that assume equilibrium (zero flow) under all conditions. Periodic boundary conditions are widely used in simulations. Such conditions take a box of material and replicate it identically, so the potential at the corresponding edges of the box are identical. If the potentials are identical, current will not flow. Periodic boundary conditions of this sort are incompatible with current flow from one boundary to the other. Voltage clamp experiments, and natural biological function involve current flow from one boundary to another.

A hybrid approach is needed, starting with the essential knowledge of structure, but computing only those parts of the structure used by biology to sense voltage. In close packed ('condensed') systems like the voltage sensor, or ionic solutions, '*everything interacts with everything else*' because electric fields are long ranged as well as exceedingly strong [14]. In ionic solutions, ion channels, even in enzyme active sites, steric interactions that prevent the overfilling of space in well-defined protein structures are also of great importance as they produce long range correlations [17]. Closely packed charged systems are best handled mathematically by energy variational methods. Energy variational methods guarantee that all variables satisfy all equations (and boundary conditions) at all times and under all conditions and are thus always consistent. We have used the energy variational approach developed in [12, 20] to derive a consistent model of gating charge movement, based on the basic features of the structure of crystallized channels and voltage sensors. The schematic of the model is shown below. The continuum model we use simulates the mechanical dynamics in a single voltage sensor, while the experimental data from [7] is from many independent voltage sensors. Ensemble averages of independent voltage sensors recording is roughly equivalent to macroscopic continuum modeling in a single voltage sensor. It is unlikely one can do better until experiments determine the structural source of the variance in the voltage sensor and its time, voltage, and chemical dependence on ion type and concentration.

---

[1] Permanent charge is our name for a charge or charge density independent of the local electric field, for example, the charge and charge distribution of $Na^+$ but not the charge in a highly polarizable anion like $Br^-$ or the nonuniform charge distribution of $H_2O$ in the liquid state with its complex time dependent (and perhaps nonlinear) polarization response to the local electric field.

**3. Mathematical model**

The reduced mechanical model for a voltage sensor is shown in Fig. 1 with four arginines $R_i$, $i$=1, 2, 3, 4, each attached to S4 helix by identical spring with spring constant $K$. The electric field will drag these four arginines since each arginine carries +1 charge. In addition, the charged arginines move the group itself. S4 connect to S3 and S5 at its two ends by identical springs with spring constant $K_{S4}/2$. Once the membrane is depolarized from say −90mV inside negative, to +10mV inside positive, arginines together with S4 will be driven towards extracellular side. When the membrane is repolarized from say +10mV volts to −90mV, the arginines move to the intracellular side. This movement is the basic voltage sensing mechanism of ion channel. The movement of S4 triggers the opening or closing of the lower gate mainly consisting of S6 of the permeant ion channel by a mechanism widely assumed to be mechanical, although we hasten to say that electrical aspects of the linker motion are likely to be involved, as well, perhaps acting as a deterministic trigger and control for the otherwise stochastic sudden discontinuous change of current seen in single channel measurements of ion flow through the permeation channel.

When arginines are driven by electric field, they are forced to move through a hydrophobic plug, composed by several nonpolar amino acids from S1, S2 and S3 [25]. Arginines moving from vestibule, a hydrated lumen, on one side, though the hydrophobic plug, and entering into the vestibule on another side would involve dehydration when passing through the hydrophobic plug. Therefore, arginines will encounter a barrier in the potential of mean force, an energy barrier, mainly dominated by the solvation energy [42]. Note that $Na^+$ and $Cl^-$ (used here as the only solution ions, for simplicity) can only exist in vestibules and are not allowed into the hydrophobic plug. The bottoms of the two vestibules on each side of the system act as impermeable walls for $Na^+$ and $Cl^-$. When the voltage is turned on and off, these two walls will form a pair of capacitors storing/releasing net ion charges in their electric double layers (EDL).

Molecular dynamics considers every particle individually and must deal with its enormous thermal motion (more or less at the speed of sound 1,000 m/s or 1 nm per nanosecond), as it seeks to compute gating currents on the biological time scale. We use a continuum treatment that deals with the thermal motion on the average, as temperature or diffusion. Such methods often include the correlations of the average electric field, e.g., Poisson-Nernst-Planck (PNP), but not those that differ from atom to atom, as a glance at the structure of any protein shows is important. Here we include a steric effect in the PNP model [20], and so deal with many of the atomic scale correlations ignored in classical PNP. We hasten to add that it is not at all clear how well molecular dynamics deals with correlations produced by the average electric field, particularly when they include nonequilibrium components that drive current flow. Classical molecular dynamics assumes equilibrium and uses periodic boundary conditions that have certain difficulties dealing with the current flow and nonperiodic electric fields found in real devices and channels, as we have previously mentioned. Continuum models can adequately describe the time behavior of a voltage sensor in action, in which the time span is on the order of several hundreds of milliseconds.

The four arginines $R_i$, $i$=1, 2, 3, 4, are described by their individual density distributions (concentrations) $c_i$, $i$=1, 2, 3, 4, allowing the arginines to interact with Na⁺ and Cl⁻ in vestibules. The density (concentration) distributions represent probability density functions as shown explicitly in the theory of stochastic processes, used to derive such equations in [29] using the general methods of [28]. This practice is used widely in quantum mechanics [3, 34]. The important issue here is how well the correlations are captured in the continuum model. Some are more likely to be faithfully captured in molecular dynamics simulations (e.g., more or less local hard sphere interactions), others in continuum models (e.g., correlations induced by far field boundary conditions).

Here we treat the S4 itself as a rigid body so we can capture the basic mechanism of a voltage sensor without considering the full details of structure, which might lead to a three dimensional model hard to compute in the time available. We construct an axisymmetric 1D model with a three-zone geometric configuration illustrated in Fig. 2 following Fig. 1. Zone 1 with $z \in [0, L_R]$ is the intracellular vestibule; zone 2 with $z \in [L_R, L_R + L]$ is the hydrophobic plug; zone 3 with $z \in [L_R + L, 2L_R + L]$ is the extracellular vestibule. Arginines, Na⁺, and Cl⁻ can all reside in zone 1 and 3. Zone 2 only allows the residence of arginines, albeit with a severe hydrophobic penalty because of their permanent charge, in a region of low dielectric coefficient hence called hydrophobic. Spatial variation of dielectric coefficient allows construction of many useful devices [15, 41]. Types of boundary conditions and relevant sizes of this geometric configuration are also specified in Fig. 2. Note the no-flux boundary conditions specified in Fig. 2. One prevents Na⁺ and Cl⁻ from entering the hydrophobic plug (zone 2) with low dielectric coefficient. The other boundary condition constrains S4 motion and so prevents the arginines from leaving the vestibules into intracellular/extracellular domains.

We first non-dimensionalize all physical quantities as follows,

$$\tilde{c}_i = \frac{c_i}{c_0}, \ \tilde{\phi} = \frac{\phi}{k_B T/e}, \ \tilde{U} = \frac{U}{k_B T}, \ \tilde{s} = \frac{s}{R}, \ \tilde{t} = \frac{t}{R^2/D_x}, \ \tilde{D}_i = \frac{D_i}{D_x}, \ \tilde{g}_{ij} = \frac{g_{ij}}{k_B T/c_0}, \ \tilde{J}_i = \frac{J_i}{c_0 D_x/R}, \ \tilde{I} = \frac{I}{ec_0 D_x R},$$

where $c_i$ is concentration of species $i$, with $i$=Na⁺, Cl⁻, 1, 2, 3, and 4. Each is scaled by $c_0$ which is the bulk concentration of NaCl at the intracellular/extracellular domains. Here $c_0$ is set to be 184mM, equal on both sides, so that the Debye length $\lambda_D = \sqrt{\frac{\varepsilon_r \varepsilon_0 k_B T}{c_0 e^2}}$ is 1nm when the relative permittivity $\varepsilon_r = 80$. Note that $c_i$, $i$=1, 2, 3, and 4 needs to satisfy the following additional constraint $\int_0^{L+2LR} A(z) c_i dz = 1$ due to uniqueness of each arginine. $\phi$ is the electric potential scaled by $k_B T/e$ with $k_B$ being the Boltzmann constant; $T$ the temperature; $e$ the elementary charge. All relevant external potentials $U$ are scaled by $k_B T$. All sizes $s$ are scaled by $R$, which is the radius of vestibule as shown in Fig. 2. $R$=1nm here. The time $t$ is scaled by $R^2/D_x$, with $D_x$ being a diffusion coefficient that can be adjusted later to be consistent with the time spans of on/off currents measured in experiments (caused by the movement of arginines). The diffusion coefficient of species *i* is scaled by $D_x$. The coupling constant $g_{ij}$ of PNP-steric model based on combining rules of Lennard Jones, representing the strength of steric interaction between species *i* and *j*, is scaled by $k_B T/c_0$ [20]. Note that here we only consider steric interaction among arginines, which is the most significant since arginines are generally crowded in hydrophobic plug and vestibules. For simplicity, we assume $g_{ij} = \begin{cases} g, \forall i \neq j \\ 0, \forall i = j \end{cases}$, $i,j = 1,2,3,4$. The flux density of species *i*, $J_i$, is scaled by $c_0 D_x / R$, and therefore the electric current *I* is scaled by $e c_0 D_x R$. For simplicity of notation, we will drop ~ for all dimensionless quantities from here on.

Based on Fig. 2, the governing 1D dimensionless PNP-steric equations are expressed below. The first one is Poisson equation:

$$-\frac{1}{A}\frac{d}{dz}\left(\Gamma A \frac{d\phi}{dz}\right) = \sum_{i=1}^{N} z_i c_i, \qquad i = \text{Na, Cl, 1, 2, 3, 4}, \tag{1}$$

with $z_{Na} = 1$, $z_{Cl} = -1$, $z_i = z_{arg} = 1, i = 1,2,3,4$, $\Gamma = \frac{\lambda_D^2}{R^2}$ and $A(z)$ being the cross-sectional area. For zones 1 and 3, $\Gamma = 1$ since here the arginines are fully hydrated with $\varepsilon_r = 80$. For zone 2, we assume a hydrophobic environment *with* $\varepsilon_r = 8$, and therefore $\Gamma = 0.1$.

The second equation is the species transport equation based on conservation law:

$$\frac{\partial c_i}{\partial t} + \frac{1}{A}\frac{\partial}{\partial z}(A J_i) = 0, \qquad i = \text{Na, Cl, 1, 2, 3, 4}. \tag{2}$$

with the content of $J_i$ expressed below based on Nernst-Planck equation for Na⁺ and Cl⁻:

$$J_i = -D_i \left(\frac{\partial c_i}{\partial z} + c_i z_i \frac{\partial \phi}{\partial z}\right), \qquad i = \text{Na, Cl}, \quad z \in \text{zones 1 and 3}, \tag{3}$$

and based on Nernst-Planck equation with steric effect and some imposed potentials for 4 arginines $c_i$, *i*=1, 2, 3 and 4, based on Fig. 2,

$$J_1 = -D_1 \left(\frac{\partial C_1}{\partial z} + z_{arg} C_1 \frac{\partial \phi}{\partial z} + C_1 \left(\frac{\partial V_1}{\partial z} + \frac{\partial V_b}{\partial z}\right) + g C_1 \left(\frac{\partial C_2}{\partial z} + \frac{\partial C_3}{\partial z} + \frac{\partial C_4}{\partial z}\right)\right), \quad z \in \text{all zones}, \tag{4}$$

$$J_2 = -D_2 \left(\frac{\partial C_2}{\partial z} + z_{arg} C_2 \frac{\partial \phi}{\partial z} + C_2 \left(\frac{\partial V_2}{\partial z} + \frac{\partial V_b}{\partial z}\right) + g C_2 \left(\frac{\partial C_1}{\partial z} + \frac{\partial C_3}{\partial z} + \frac{\partial C_4}{\partial z}\right)\right), \quad z \in \text{all zones}, \tag{5}$$

$$J_3 = -D_3 \left(\frac{\partial C_3}{\partial z} + z_{arg} C_3 \frac{\partial \phi}{\partial z} + C_3 \left(\frac{\partial V_3}{\partial z} + \frac{\partial V_b}{\partial z}\right) + g C_3 \left(\frac{\partial C_1}{\partial z} + \frac{\partial C_2}{\partial z} + \frac{\partial C_4}{\partial z}\right)\right), \quad z \in \text{all zones}, \tag{6}$$

$$J_4 = -D_4 \left(\frac{\partial C_4}{\partial z} + z_{arg} C_4 \frac{\partial \phi}{\partial z} + C_4 \left(\frac{\partial V_4}{\partial z} + \frac{\partial V_b}{\partial z}\right) + g C_4 \left(\frac{\partial C_1}{\partial z} + \frac{\partial C_2}{\partial z} + \frac{\partial C_3}{\partial z}\right)\right), \quad z \in \text{all zones}. \tag{7}$$

The first and second terms in Eqs. (3-7) describe diffusion and electro-migration term respectively. The third term in Eqs. (4-7) are external potential terms with $V_i$, *i*=1, 2, 3 and 4 being the constraint potential for the 4 arginines $c_i$ to S4 represented here by a spring connecting each arginine $c_i$ to S4 as shown in Fig. 1. It is expressed as

$$V_i(z,t) = K(z - (z_i + Z_{S4}(t)))^2, \tag{8}$$

where *K* is the spring constant, $z_i$ is the fixed anchoring position of the spring for each arginine $c_i$ on S4, $Z_{S4}(t)$ is the center-of-mass *z* position of S4 by treating S4 as a rigid body. Here we set $z_1$=0.6, $z_2$=0.2, $z_3$=-0.2, $z_4$=-0.6 given from the structure that gives the arginine anchoring interval on S4 as 0.4nm. $Z_{S4}(t)$ follows the motion of equation based on spring-mass system:

$$m_{S4}\frac{d^2Z_{S4}}{dt^2} + b_{S4}\frac{dZ_{S4}}{dt} + K_{S4}(Z_{S4} - Z_{S4,0}) = \sum_{i=1}^{4} K\left(z_{i,CM} - (z_i + Z_{S4})\right), \qquad (9)$$

where $m_{S4}$, $b_{S4}$ and $K_{S4}$ are mass, damping coefficient and restraining spring constant for S4. $Z_{S4,0}$ is the natural position of $Z_{S4}(t)$ when the net force on the right hand side of Eq. (9) vanishes. Here, $z_{i,CM}$ is the center of mass for the set of arginines $c_i$, which can be calculated by

$$z_{i,CM} = \frac{\int_0^{L+2LR} A(z) z c_i dz}{\int_0^{L+2LR} A(z) c_i dz}, \quad i=1, 2, 3, 4. \qquad (10)$$

We assume that the spring mass system for S4 is over-damped, which means the inertia term in Eq. (9) can be neglected.

Another external potential in the third term of Eqs. (4-7) is the energy barrier $V_b$, which is nonzero only in zone 2 representing mainly the solvation energy barrier from hydrophobicity. Generally, the accurate shape of $V_b$ in zone 2 requires a calculation of potential mean force (PMF) using molecular dynamics simulations by the structure of protein and the possible change in the PMF with membrane potential, ionic conditions, consistent with the distribution of charge and boundary conditions, and so on. However, here we simply assume a hump shape for the PMF, while we await proper calibrated calculations of the PMF and its dependence on membrane potential, etc.

$$\begin{cases} V_{b,max}\left(\tanh(5(z-L_R)) - \tanh(5(z-L-L_R)) - 1\right), & z \in \text{zone 2}, \\ 0, & z \in \text{zones 1 and 3}. \end{cases} \qquad (11)$$

with $V_{b,max}$ set to be 5 in the current situation [42]. If we set $V_{b,max}$ too large, the gating current would be very small since it would be very difficult for arginines to move across this barrier. Note that the tanh function is designated to smooth the otherwise top-hat-shape barrier profile, with its awkward infinite slopes not good for (numerical) differentiation. It is also generally believed that energy barrier in a protein structure does not have a jump. The last term in Eqs. (4-7) is the steric term that accounts for steric interaction among arginines [20]. Here we set $g$=0.5. Larger $g$ implies larger steric effect, but $g$ cannot be arbitrarily large due to the limitation of stability.

Governing equations Eqs. (1-7) were derived by energy variational method based on the following energy (in dimensional form):

$$E = \int_V \left[ k_B T \sum_{all\ i} c_i \log c_i - \frac{\varepsilon_0 \varepsilon_r}{2} |\nabla \phi|^2 + \sum_{all\ i} z_i e\, c_i \phi + \sum_{arginines} (V_i + V_b) c_i \right. \\ \left. + \sum_{arginines\ i,j} \frac{g_{ij}}{2} c_i c_j \right] dV,$$

where the first term is entropy; second and third terms are electrostatic energy; fourth term is constraint and barrier potential for arginines; last term is the steric energy term based on Lennard-Jones potential [12, 20]. The Poisson equation Eq. (1) is derived by

$$\frac{\delta E}{\delta \phi} = 0,$$

and species flux densities in Eqs. (3-7) are derived by

$$\mu_i = \frac{\delta E}{\delta c_i}, \quad J_i = -\frac{D_i}{k_B T} c_i \nabla \mu_i,$$

where $\mu_i$ is the chemical potential of species $i$.

Also, here we assume quasi-steady state for Na$^+$ and Cl$^-$, which means $\frac{\partial c_i}{\partial t} = 0, i = \text{Na, Cl}$. This is justified by the fact that the diffusion coefficients of Na⁺ and Cl⁻ in vestibules are much larger than the diffusion coefficient of arginine based on the very narrow time span of the leading spike of gating current measured in experiments. The spike comes from the linear capacitive current of vestibule when the command potential suddenly rises or drops. This quasi-steady state assumption is essential. Otherwise using realistic diffusion coefficients for Na⁺ and Cl⁻ would render Eqs. (1-7) too stiff to integrate in time. The spike contaminating the gating current can be removed by a simple technique called P/n leak subtraction in experiments (see Section 5.4). How to do this computationally will be discussed in section 5.3. Boundary and interface conditions for electric potential $\phi$ are

$$\phi(0) = V, \quad \phi(L_R^-) = \phi(L_R^+), \quad \Gamma(L_R^-)A(L_R^-)\frac{d\phi}{dz}(L_R^-) = \Gamma(L_R^+)A(L_R^+)\frac{d\phi}{dz}(L_R^+),$$

$$\phi(L_R + L^-) = \phi(L_R + L^+), \quad \Gamma(L_R + L^-)A(L_R + L^-)\frac{d\phi}{dz}(L_R + L^-) = \Gamma(L_R + L^+)A(L_R + L^+)\frac{d\phi}{dz}(L_R + L^+),$$

$$\phi(2L_R + L) = 0. \quad (12)$$

These are Dirichlet boundary conditions at both ends and continuity of electric potential and displacement at the interfaces between zones. Boundary and interface conditions for arginine are

$$J_i(0,t) = J_i(2L_R + L, t) = 0, \quad c_i(L_R^+, t) = c_i(L_R^-, t), \quad A(L_R^-)J_i(L_R^-, t) = A(L_R^+)J_i(L_R^+, t),$$

$$c_i(L_R + L^-, t) = c_i(L_R + L^+, t), \quad A(L_R + L^-)J_i(L_R + L^-, t) = A(L_R + L^+)J_i(L_R + L^+, t), \quad i = 1,2,3,4. \quad (13)$$

No-flux boundary conditions are placed at both ends of the gating pore to prevent arginines and S4 from entering intracellular/extracellular domains. The others are continuity of concentration and flux at interfaces between zones. Boundary conditions for Na⁺ and Cl⁻ are

$$c_{Na}(0,t) = c_{Cl}(0,t) = c_{Na}(2L_R + L, t) = c_{Cl}(2L_R + L, t) = 1,$$

$$J_{Na}(L_R, t) = J_{Cl}(L_R, t) = J_{Na}(L_R + L, t) = J_{Cl}(L_R + L, t) = 0. \quad (14)$$

Dirichlet boundary conditions are placed at both ends of the gating pore to describe the concentrations for Na⁺ and Cl⁻ are bulk concentration over there and no-flux boundary conditions at both ends of hydrophobic plug to describe the impermeability of Na⁺ and Cl⁻ into hydrophobic plug.

Besides the main input parameter $V$, which is the voltage bias (command potential) applied, other parameters like $D_i$, $i$=1, 2, 3, 4, $K$, $K_{S4}$, $b_{S4}$ are also required. Results are especially sensitive to the values of $K$, $K_{S4}$, $b_{S4}$. We have tried and found $D_i$=50, $i$=1,2,3,4, $K$=3, $K_{S4}$=3, $b_{S4}$=1.5 fit best with experiment [7]. We compare computational outputs with important experiments: (1) gating current versus voltage curve (IV) and (2) gating charge versus voltage curve (QV) as well as (3) gating current vs. time curves.

Usually the electric current in the ion channel is treated simply as the flux of charge and is uniform in $z$ when steady. This is not so in the present non-steady dynamic situation, since storing and releasing of charge in vestibules and the channel are involved. Here the flux of charge at the middle of hydrophobic plug, $z= L_R+L/2$, was computed to estimate the experimentally observed gating current. However, it is actually impossible to experimentally measure the current within the channel since it is impossible to put a probe in the middle of the hydrophobic plug and optical methods to record these variables are so far unavailable. In experiments, the voltage clamp technique is used, and on/off gating current through the membrane is measured in experiments, which should be equal to the flux of charge at $z$=0 in the present frame work as shown in Fig. 2. The flux of charges at any $z$ position $I(z,t)$ can be related to the flux of charges at $z$=0, $I(0,t)$, simply by charge conservation:

$$\frac{\partial}{\partial t}Q_{net}(z,t) = I(0,t) - I(z,t), \quad (15)$$

where

$$Q_{net}(z,t) = \int_0^z A(\xi) \sum_{\text{all } i} z_i c_i d\xi, \quad (16)$$

and flux of charges at any z position $I(z,t)$ is defined by
$$I(z,t) = A(z) \sum_{\text{all } i} z_i J_i(z,t). \quad (17)$$

We may as well identify $\frac{\partial}{\partial t} Q_{net}(z,t)$ as the displacement current, and denote it as $I_{disp}(z,t)$, since it will be shown later that Eq. (15) is equivalent to Ampere's law in Maxwell's equations, and $\frac{\partial}{\partial t} Q_{net}(z,t)$ is exactly the displacement current in Ampere's law. A general discussion can be found in [10]. Hence, Eq. (15) can be simply re-written as
$$I_{tot}(z,t) = I(z,t) + I_{disp}(z,t) = I(0,t), \quad (18)$$
where we define the sum of displacement current and flux of charges as the total current $I_{tot}(z,t)$. The z-distribution of the total current should be uniform by Kirchhoff's law, which will be computationally verified in section 5.3. We are also interested in observing the net charge at vestibules. Taking net charge at intracellular vestibule, $Q_{net}(L_R, t)$, as an example, the net charge consists of arginine charges and their counter charges formed by the EDL of ionic solution over there. Generally, electro-neutrality is approximate but will not be exact there. Note that the existence of EDL's at vestibules, acting as a pair of capacitors, is an unavoidable consequence of discontinuities in constitutive properties like permanent charge density and dielectric coefficients. Realistic atomic scale simulations will have EDL and capacitive currents. More about flux of charge, displacement current and net charge at vestibules will be discussed in section 5.3.

Eq. (15) is consistent with Ampere's law in Maxwell's equations:
$$\nabla \times \left(\frac{\vec{B}}{\mu_0}\right) = \varepsilon_0 \varepsilon_r \frac{\partial \vec{E}}{\partial t} + \vec{J}, \quad (19)$$
or equivalently,
$$\nabla \cdot \left(\varepsilon_0 \varepsilon_r \frac{\partial \vec{E}}{\partial t} + \vec{J}\right) = 0, \quad (20)$$
where $\vec{E}$ is the electric field and $\vec{J}$ is flux density of charge (current density). Eq. (20) tells us that the current is conserved everywhere and it consists of flux of charges $\vec{J}$ and displacement current $\varepsilon_0 \varepsilon_r \frac{\partial \vec{E}}{\partial t}$. Eq. (20) can be derived by Poisson equation and species transport equation like Eq. (1) and Eq. (2). Starting from Poisson equation in dimensional form:
$$-\nabla \cdot (\varepsilon_0 \varepsilon_r \nabla \phi) = \rho + \sum_i z_i e c_i, \quad (21)$$
or equivalently
$$\nabla \cdot (\varepsilon_0 \varepsilon_r \vec{E}) = \rho + \sum_i z_i e c_i. \quad (22)$$
Taking time derivative of Eq. (22),
$$\nabla \cdot \left(\varepsilon_0 \varepsilon_r \frac{\partial \vec{E}}{\partial t}\right) = \sum_i z_i e \frac{\partial c_i}{\partial t}, \quad (23)$$
and using species transport equation based on mass conservation,
$$\frac{\partial c_i}{\partial t} + \nabla \cdot \vec{J}_i = 0, \quad (24)$$
then

$$\nabla \cdot \left(\varepsilon_0 \varepsilon_r \frac{\partial \vec{E}}{\partial t}\right) = \sum_i z_i e \frac{\partial c_i}{\partial t} = -\nabla \cdot \sum_i z_i e \vec{J}_i = -\nabla \cdot \vec{J}, \qquad (25)$$

which is exactly Eq. (20) by defining

$$\vec{J} = \sum_i z_i e \vec{J}_i. \qquad (26)$$

Casting Eq. (20) into the present 1D framework and integrate in space from 0 to z, we have

$$\varepsilon_0 \varepsilon_r A(z) \frac{\partial E(z,t)}{\partial t} + I(z,t) = \varepsilon_0 \varepsilon_r A(0) \frac{\partial E(0,t)}{\partial t} + I(0,t). \qquad (27)$$

Comparing with Eq. (18),

$$\varepsilon_0 \varepsilon_r A(z) \frac{\partial E(z,t)}{\partial t} - \varepsilon_0 \varepsilon_r A(0) \frac{\partial E(0,t)}{\partial t} = I_{disp}(z,t), \qquad (28)$$

which justifies the naming of displacement current in Eq. (18).

To construct the QV curve, we calculate $Q_1 = \int_0^{L_R} A(z) \sum_{i=1}^4 c_i dz$, $Q_2 = \int_{L_R}^{L_R+L} A(z) \sum_{i=1}^4 c_i dz$, $Q_3 = \int_{L_R+L}^{2L_R+L} A(z) \sum_{i=1}^4 c_i dz$, which are the amounts of arginine found in zone 1, 2 and 3, respectively. Usually $Q_2 \approx 0$ due to the energy barrier $V_b$ in zone 2. Arginines tend to jump across zone 2 when driven from zone 1 to zone 3 when the voltage $V$ is turned on. The number of arginines that move and settles at zone 3 depends on the magnitude of $V$. Besides IV and QV curves, the time course of the movement of the arginines and of gating charge (and $z_{i,CM}(t)$ and $Z_{S4}(t)$), are important to report here. The movement of arginines can (almost) be recorded in experiments nowadays by optical methods. Many qualitative models accounting for the movement of S4 and conformation change of the voltage sensor have been proposed. Readers are referred to the review paper [37] for more details.

### 4. Numerical method

High-order multi-block Chebyshev pseudospectral methods are used here to discretize Eqs. (1-2) in space [38]. The resultant semi-discrete system is then a set of coupled ordinary differential equations in time and algebraic equations (an ODAE system) [2]. The ordinary differential equations are chiefly from Eq. (2), and algebraic equations are chiefly from Eq. (1) and boundary/interface conditions Eq. (12-14). This system is further integrated in time by an ODAE solver (ODE15S in MATLAB [31-32]) together with appropriate initial condition. ODE15S is a variable-order-variable-step (VSVO) solver, which is highly efficient in time integration because it adjusts the time step and order of integration. High-order pseudospectral methods generally provide excellent spatial accuracy with economically practicable resolutions. A combination of these two techniques makes the whole computation very efficient. This is particularly important, since numerous computations have to be tried during the tuning of parameters.

### 5. Results and discussions

Here we explored several parameter values to obtain charge movement with kinetics and steady state properties similar to the experimentally recorded gating currents. Numerical results based on the mathematical model described above were calculated and compared with experimental measurements [7]. Our one dimensional continuum model has advantages and disadvantages. The lack of three-dimensional structural detail means of course that some details of the gating current and charge cannot be reproduced. It should be noted however that to reproduce those, one needs more than just static structural detail. One must also know how the structures (particularly their permanent and polarization charge) change after a command potential is applied, in the ionic conditions of importance. Structural detail that does not depend on time can be expected to give an incomplete, probably misleading image of mechanism (imagining studying an internal combustion engine without seeing the pistons or valves moving). The 1D model has the important advantages that it computes the actual experimental results on the actual experimental time scale, in realistic ionic solutions and with far field boundary conditions actually used in voltage clamp experiments. It also conserves current as we will discuss later

**5.1 QV curve**

When the membrane and voltage sensor is held at a large inside negative potential (e.g., hyperpolarized to inside negative say -90mV), S4 is in a down position and all arginines stay in the intracellular vestibule. When the potential is made more positive (e.g., depolarized to +10mV, inside positive), S4 is in the up position and all arginines are at the extracellular vestibule.

The voltage dependence of the charge (arginines) transferred from intracellular vestibule to extracellular vestibule is characterized as a QV curve in experimental papers and it is sigmoidal in shape [7]. Fig. 3(a) shows that our computed QV curve— the dependence of $Q_3$ on $V$—is in very good agreement with the experiment [7]. Fig. 3(b) shows the steady-state distributions of Na$^+$, Cl$^-$ and arginines in the inside negative, hyperpolarized situation ($V$=-90mV). As we can see, all arginines stay at intracellular vestibule, and none of arginines transferred to extracellular vestibule ($Q_3 \approx 0$).

Fig. 3(c) shows the situation at $V$=-48mV, which is the midpoint of QV curve. As we can see, each vestibule has half of the arginines staying in it ($Q_3 = 2$): R1 and R2 are expected in extracellular vestibule, and R3 and R4 in intracellular vestibule. The concentration distributions of arginines shown in Fig. 3(c), interpreted as individual probability density functions, show R1 and R2 residing more in extracellular vestibule, and R3 and R4 on the contrary more in intracellular vestibule. Note that there are almost no arginines in zone 2 (hydrophobic plug) due to the energy barrier in it. If the command potential at the midpoint ($Q_3 = 2$) is $V$=0mV, the unforced natural position of S4 is at the middle of gating pore, i.e., $Z_{S4,0} = L_R + 0.5L$. In experimental measurement [7], $V$ is actually -48mV instead of 0mV for the midpoint. This requires $Z_{S4,0}$ to be biased from $L_R + 0.5L$ to $Z_{S4,0} = L_R + 0.5L + 1.591$nm. Fig. 3(d) shows the situation at full depolarization ($V$=-8mV). As we can see, all arginines move to extracellular vestibule ($Q_3 \approx 4$).

**5.2 Gating current**

Fig. 4 shows the time course of gating currents, observed as flux of charge at the middle of gating pore $I(L_R + L/2, t)$, due to the movement of arginines when the membrane is largely depolarized, and partially depolarized. In the case of large depolarization, $V$ rises from -90mV to -8mV at $t$=10, holds on till $t$=150, and drops back to -90mV as shown in Fig. 4(a). Time course of gating current and contributions of individual arginines are shown in Fig. 4(b). We observe that the rising order of each current component follows the moving order of R1, R2, R3 and R4 when depolarized, and that order is reversed when repolarized later. The area under the gating current is the amount of charge moved. Since arginines move forward and backward in this depolarization/repolarization scenario, the areas under the ON current (arginines moving forward) and the OFF current (arginines moving backward) are same. The areas are equal for each component of current as well. The equality of area is an important signature of gating current that contrasts markedly with the properties of ionic current. Equality of area has in fact been used as a signature that identifies gating current and separates it from ionic current [26, 5]. Since almost all arginines move from intracellular to extracellular vestibule when the command potential is suddenly made more positive in a large depolarization, the area under each current component is then very close to 1. In the case of partial depolarization, $V$ rises from -90mV to -50mV at $t$=10, holds on till $t$=150, and drops back to -90mV as shown in Fig. 4(c). The time course of gating current and its four components contributed by each arginine for this situation is shown in Fig. 4(d). Under this partial polarization, not all arginines move past the middle of hydrophobic plug due to weaker driving force in partial polarizations compared with large depolarization case. This can be seen in Fig. 4(d), where areas under each component current are different, and based on these various areas R1 moves more towards extracellular vestibule than R2; R2 more so than R3; and etc.

The gating currents shown above can be better understood by looking at a sequence of snapshots showing the spatial distribution of electric potential, species concentration and electric current. The distributions at several times are shown in Fig. 5 for the case of sudden change in command voltage to a more positive value, a large depolarization, and Fig. 6 for the case of small positive going change in potential, a partial depolarization. Fig. 5 shows that almost all arginines are driven from intracellular to extracellular vestibules in case of large depolarization, but only part of the arginines move in case of partial depolarization (Fig. 6). The electric potential profiles at $t$=13 and $t$=148 show that the profile of electric potential as arginines move from left to right even though the voltage is maintained constant across the sensor. This is not a constant field system at all! Note that slight bulges in electric potential profile exist wherever arginines are dense. This can be easily understood by understanding the effect of Eq. (1) on a concave spatial distribution of electric potential.

In Figs. 5 and 6, total current defined in Eq. (18), though changing with time, is always constant in $z$ at all times, which is no surprise because it must satsify Kirchhoff's law. At $t$=13, a time that gating current is not quiescent as shown in Fig. 4(b) and 4(d), we can particularly visualize the $z$-distributions of flux of charges $I(z,t)$, displacement of current $I_{disp}(z,t)$ and total current $I_{tot}(z,t)$ individually in Figs. 5 and 6.

### 5.3 Flux of charges at different locations

Flux of charges $I(z,t)$, together with displacement of current $I_{disp}(z,t)$ and total current $I_{tot}(z,t)$, depicted in Figs. 5 and 6 deserves more discussions here. Flux of charges at the middle of gating pore, $I(L_R + L/2, t)$, and both ends of gating pore, $I(0,t)$ and $I(2L_R + L, t)$, should be computed by

$$I\left(L_R + \frac{L}{2}, t\right) = A\left(L_R + \frac{L}{2}\right) \sum_{arginines} z_i J_i \left(L_R + \frac{L}{2}, t\right), \quad (29)$$

$$I(0,t) = A(0) \sum_{i=N,Cl} z_i J_i(0,t), \quad (30)$$

$$I(2L_R + L, t) = A(2L_R + L) \sum_{i=Na,Cl} z_i J_i(2L_R + L, t). \quad (31)$$

Except $I\left(L_R + \frac{L}{2}, t\right)$, $I(0,t)$ and $I(2L_R + L, t)$ are trivially zero due to the implement of quasi-steadiness $\frac{\partial c_i}{\partial t} = 0, i =$ Na, Cl, in vestibules, which causes $J_{Na}$ and $J_{Cl}$ to be uniform in vestibules by Eq. (2), and further become zero by the no-flux boundary conditions for $Na^+$ and $Cl^-$ at the bottom of vestibules as described in Eq. (14). We have to alternatively reconstruct $I(0,t)$ and $I(2L_R + L, t)$ by charge conservation of $Na^+$ and $Cl^-$,

$$I(0,t) = \frac{d}{dt}\int_0^L A(z) \sum_{\text{Na,Cl}} z_i c_i dz, \tag{32}$$

$$I(2L_R + L, t) = -\frac{d}{dt}\int_{L+L_R}^{L+2L_R} A(z) \sum_{\text{Na,Cl}} z_i c_i dz. \tag{33}$$

After obtaining $I(0,t)$ and $I(2L_R + L, t)$, we can further reconstruct flux of charges $I(z,t)$ at zone 1 (intracellular vestibule) and zone 3 (extracellular vestibule) by charge conservation of $Na^+$ and $Cl^-$ again,

$$I(z,t) = I(0,t) - \frac{d}{dt}\int_0^z A(z) \sum_{\text{Na,Cl}} z_i c_i dz, \quad z \in [0, L_R], \tag{34}$$

$$I(z,t) = I(2L_R + L, t) + \frac{d}{dt}\int_z^{2L_R+L} A(z) \sum_{\text{Na,Cl}} z_i c_i dz, \quad z \in [L_R + L, 2L_R + L]. \tag{35}$$

Once flux of charges is analyzed, we can then compute the displacement current based on finding the time derivative of Eq. (16). Finally we sum up flux of charges $I(z,t)$ and displacement of current $I_{disp}(z,t)$ to obtain total current $I_{tot}(z,t)$ and verify that it satisfies Eq. (18) and is uniform in $z$. This verification is shown in Figs. 5 and 6 at several various times, and is fact true at any time as well.

In the bottom rows of Figs. 5 and 6 at $t=13$, we observe that $I(z,t)$ is generally non-uniform in $z$ and is accompanied by congestion/decongestion of arginines in between. However $I(z,t)$ is almost uniform at zone 2 (hydrophobic plug), which means almost no congestion/decongestion of arginines occurs there, and therefore no contribution to the displacement current $\frac{d}{dt}Q_{net}(z,t)$ from zone 2. This is no surprise since arginines can hardly reside at zone 2 due to energy barrier in it. Because of the energy barriers, once arginines leave zone 1 (intracellular vestibule), they immediately jump across zone 2 and enter into zone 3 (extracellular vestibule).

Several things are worth noting in the time courses of $I\left(L_R + \frac{L}{2}, t\right)$, and $I(0, t)$ illustrated in Fig. 7(a) under the case of large depolarization. First, $I\left(L_R + \frac{L}{2}, t\right)$ is noticeably larger than $I(0,t)$ in ON period. This is because their difference, exactly the displacement current $I_{disp}$, is always negative at zone 2 when depolarized, since arginines are leaving zone 1 and make $\frac{d}{dt}Q_{net} < 0$ for zone 2. It is expected the area under the time course of $I\left(L_R + \frac{L}{2}, t\right)$ would be very close to 4e, as verified by the time courses of $Q_3$ in Fig. 7(b), while the counterpart area of experimentally measurable $I(0,t)$ would be less than 4e due to its smaller magnitude compared with $I\left(L_R + \frac{L}{2}, t\right)$. This may account for some experimental observations that at most 13e [4, 27, 30], instead of 16e, are moved during full depolarization in 4 voltage sensors (for a single ion channel) based on calculating the area under voltage-clamp gating current. Therefore, flux of charge at any location of zone 2, though impossible to measure in experiments so far, will give us amount of arginines moved during depolarization more reliably than the measurable $I(0,t)$.

Second, we see in Fig. 7(a) with magnification in its inset plot that, as in experiments, $I(0,t)$, but not $I\left(L_R + \frac{L}{2}, t\right)$, has contaminating leading spikes in ON and OFF parts of the current. These spikes are linear capacitive currents from solution EDL of vestibules caused by sudden rising and dropping of command potential, and are called the early transient gating current in experiments [33, 35-36]. In voltage-clamp experiments, subtracting this linear capacitive component and removing the spike from gating current is done by 'leak subtraction', in various forms, e.g., P/4 (see Section 5.4) In reality, this linear capacitive current that is subtracted in this procedure comes from both the lipid bilayer membrane in parallel with the channel and vestibule solution in series with the channel. Here, we only considered capacitive current from solution EDL of vestibule and ignored the capacitive current of the membrane in parallel with the channel (which is actually much larger than vestibule capacitive current) because we use Dirichlet boundary conditions for $\phi$ at both ends of gating pore in Eq. (12). Following the idea of experiment [7], we calculated $I(0,t)$ with $V$ rising from -150mV to -140mV at $t$=10, holding on till $t$=150, and dropping back to -150mV. The reason to choose the rising range of $V$ from -150mV to -140mV is that essentially none of the arginines move in this hyperpolarized situation. The voltage step only quickly charges and discharges solution EDL in vestibules, and the computational time course of $I(0,t)$ is just two spikes when the command potential rises suddenly and drops suddenly later. Subtracting this hyperpolarized $I(0,t)$, multiplied by a proportion factor, from its original counterpart will then remove the spikes, and the unspiked $I(0,t)$ is shown in Fig. 7(a).

Third, in Fig. 7(b), the time courses of $Q_1$ and $Q_3$ show that arginines move from the intracellular vestibule to the extracellular one when depolarized and move back to intracellular vestibule again when repolarized later. As arginines move from one vestibule to another, the concentrations of Na$^+$ and Cl$^-$ also correspondingly change with time at the vestibules, form counter charges through EDL, and balance arginine charges at vestibules. However, these movement only maintain approximate, not exact, electroneutrality as shown in Fig. 7(b). The violation of electroneutrality is produced by the displacement current, that is not negligible we note.

As mentioned above, here we used flux of charges at the middle of hydrophobic plug, $I\left(L_R + \frac{L}{2}, t\right)$, instead of experimentally measurable $I(0,t)$ to represent 'the gating current' in discussions. This 'gating current' just leaves out the displacement current $I_{disp}\left(L_R + \frac{L}{2}, t\right)$. We use this definition of 'gating current' for several reasons: (1) the area under time course of $I\left(L_R + \frac{L}{2}, t\right)$ gives us the amount of arginines moved during depolarization more faithfully than $I(0,t)$ as explained above. The fluxes of charge for each arginine shown in Fig. 4 (b) and (d) carry important information about how each arginine is moved by the electric field, that will be illustrated in Fig. 8. All these will not be easy to display and comprehend if we use $I(0,t)$ instead. (2) Using $I(0,t)$ as a definition of gating current would require a decontamination by removing the leading spikes in it, which is computationally costly by the procedure described above. Especially, it would pose a heavy numerical burden when doing parameter fitting, where numerous repeated computations need to be conducted; (3) The shape of $I\left(L_R + \frac{L}{2}, t\right)$ is actually close to the unspiked $I(0,t)$. Hence here we used the computationally handy $I\left(L_R + \frac{L}{2}, t\right)$ to replace $I(0, t)$ as the gating current.

## 5.4 Comparison with experimental records.

Our computations have limited fidelity at short times because of time step limitation in integrating stiff systems. The spike artifacts are one example, described previously. Experimental measurements [16, 33] of the fast transient gating current are fascinating but our calculations must be extended to deal with them.

A more general consideration is the subtraction procedure used in experiments to isolate gating current from currents arising from other sources. Channels and their voltage sensors are always accompanied by large amounts of lipid membranes. Currents through channels and in voltage sensors are 'in parallel' with large capacitive currents through lipid membranes. The lipid membrane of cells introduces a large capacitance ( $C_{lipid} \cong 6 \times 10^{-7}$ farads/cm$^2$ ) that has nothing to do with the capacitive (i.e., displacement) currents produced by charge movement in the voltage sensor. Fortunately, this capacitance $C_{lipid}$ is a nearly ideal circuit element. Current through the membrane capacitance is entirely a displacement current accurately described by $i_{cap} = C_{lipid} \frac{\partial V}{\partial t}$ with a single constant $C_{lipid}$. $V$ is the voltage across the lipid capacitor. Note that $i_{cap}$ does not include any current carried by the translocation of ions across the lipid.

In experimental measurements, this displacement current $i_{cap}$ is always present, in large amounts, because voltage sensors (and channels) are always embedded in lipid membranes. Experimental measurements mix the displacement current of the lipid membrane and the displacement current of the voltage sensor. Indeed, the lipid membrane current dominates the measurement of displacement current in native preparations and remains large in systems mutated to have unnaturally large numbers of voltage sensors.

A procedure to remove the lipid membrane current is needed if the gating current of the voltage sensor is to be measured. The procedure introduced by Schneider and Chandler [26] has been used ever since in the improved P/4 version developed by Armstrong and Bezanilla [1] reviewed and discussed in [8]. Also, see another approach in [6, 13]. Schneider and Chandler's procedure [26] estimates the so-called linear current $i_x = C_x \frac{\partial V}{\partial t}$ in conditions in which the voltage sensor and $C_x$ behaves as an ideal circuit element. An ideal capacitor has a capacitance $C_x$ independent of voltage, time, current, or ionic composition. The Schneider procedure then subtracts that linear current $i_x$ from the total current measured in conditions in which the voltage sensor does not behave as an ideal capacitor. The leftover estimates the nonlinear properties of the charge movement in the voltage sensor. That is to say it estimates the charge movement of the voltage sensor that is not proportional to the size of the voltage step used in the measurement. The leftover is called gating current here and in experimental papers.

The gating current reported in experiments however is missing a component of the displacement current of the voltage sensor, because of the Schneider and Chandler procedure that estimates $i_x$. The Schneider and Chandler procedure does not measure $i_{cap}$ by itself. The linear current $i_x$ estimated by the Schneider and Chandler procedure has more in it than the current through the lipid membrane capacitor $i_{cap}$. Rather, it contains the lipid membrane current $i_{cap}$ plus current through any structures in the membrane ('in parallel') in which current follows the law $i_x = C_x \frac{\partial V}{\partial t}$.

Clearly, some of the current produced by movements of the arginines in the voltage sensor will be a linear displacement current, a linear component of gating current. But that component is likely to be included in $i_x$ and so would not be present in the reported gating current. The linear component of gating current is removed by the experimental procedure that is needed to remove the lipid membrane displacement current $i_{cap}$. Other systems may contribute to the linear displacement current as well, for example, (1) all sorts of experimental and instrumentation artifacts and (2) displacement current in the conduction channel itself. The conduction channel of field effect transistors produces a large capacitance described by drift diffusion equations quite similar to the PNP equations of the open conduction channel.

Our procedure in the computations reported here subtracts a hyperpolarized current with arginines not moving in this situation, thereby removing all the currents carried by arginine movement, linear and nonlinear in voltage. Most systems have substantial motions that are linear in voltage (even if the system is labelled 'nonlinear'). The response of most systems to a voltage step can be written as a sum of terms, each with a different dependence on the size of the applied step of voltage V. The first term in that sum changes linearly if the voltage step is changed. Higher order terms exhibit a nonlinear dependence on V. The linear term is present in most systems, just as it is present in most Taylor expansions of nonlinear functions.

The linear component missed in experiments, and removed in these calculations, may have functional and structural significance. The voltage sensor works by sensing voltage, for example, by producing a motion of arginines. That motion—the response of the voltage sensor in this model—includes a linear component. The signal passed to the conduction channel, to control gating, is likely to include the linear component of sensor function, including the linear component of the motion of the arginines. Confusion will result if the linear component is ignored when a model is created that links the voltage sensor to the gating process of the conduction channel.

Direct measurements of the movement of arginines (e.g., with optical methods) are likely to include the linear component and so should not agree with experimental measurements of gating current or with the currents reported here.

## 5.5 Time course of arginine and S4 translocation

Time course of $Q$ (amount of arginines moved to extracellular vestibule, equal to $Q_3$ here) and center-of-mass trajectories of individual arginines ($z_{i,CM}$, $i$=1, 2, 3, 4) and S4 segment ($Z_{S4}$) are shown in Fig. 8 with (a) and (b) for the case of large depolarization and (c) and (d) for the case of partial depolarization. In the case of large depolarization, from Fig. 8(b), we can see arginines and S4 are driven towards extracellular vestibule once the electric field is turned on. Their z-positions quickly reach individual steady states with almost all arginines transferred to extracellular vestibule as previously shown in Fig. 5 and therefore $Q$ is close to 4 as shown in Fig. 8(a). Arginines and S4 move back to intracellular vestibule once the voltage drops back to -90mV. From Fig. 8(b), the forward moving order of arginines is R1, R2, R3 and R4, and the backward moving order is on the opposite R4, R3, R2 and R1. This obviously coincides with the structure. Note that S4 is initially farthest to the right but lags behind R1 and R2 during movement in depolarization as shown in Fig. 8(b). This is certainly because S4 is relaxed to an almost unforced situation close to its natural position $Z_{S4,0}$. We can further calculate the displacements of each arginine and S4 during this full depolarization, and find $\Delta z_{1,CM} \approx \Delta z_{2,CM} \approx \Delta z_{3,CM} \approx 1.93$nm, $\Delta z_{4,CM}$=1.76nm, $\Delta Z_{S4}$=1.51nm. Besides almost the same displacements for R1, R2 and R3, their average moving velocities are also very close to each other. This seems to suggest a synchronized movement among R1, R2 and R3. Also, we can see the movements of arginines contribute significantly to the movement of S4 segment. This can be seen by steady state *z*-position of S4 derived from Eq. (9),

$$Z_{S4} = \frac{K}{K_{S4}+4K}\sum_{i=1}^{4}(z_{i,CM} - z_i) + \frac{K_{S4}}{K_{S4}+4K}Z_{S4,0} = \frac{1}{5}\left[Z_{S4,0} + \sum_{i=1}^{4} z_{i,CM}\right].$$

Experimental estimates of S4 displacement during full depolarization ranged from 2-20 Å [24, 37]. This large deviation accommodates several models for voltage sensing: transporter model, helical screw model and paddle model [37]. Our $\Delta Z_{S4}$=1.51nm here is large enough that seems to agree better with experimental estimates requiring large displacements, such as the paddle model. However, our model is more based on helical screw model, which is known to have shorter displacements. A plausible explanation for our over-estimated $\Delta Z_{S4}$ is that our 1D model has made the known helical path into a linear path. It would be shorter for displacement perpendicular to the plane of the membrane. The diameter of an alpha helix between alpha carbons is approximately 1 nm, therefore if the rotation were 90°, the displacement in 1D would be extended by 0.5×0.5×π=0.78 nm, making the linear translation perpendicular to the membrane only 0.73 nm. This number is very close to the displacement perpendicular to the membrane that is estimated when comparing the open-relaxed state crystal structure of Kv1.2 [9] and the consensus closed structure that has been derived from experimental measurements [40].

For partial depolarization, arginines and S4 are the same driven towards extracellular vestibule once the electric field is turned on and return when voltage drops. However, since the driving force is weaker than in a large depolarization, their *z*-positions do not reach steady states as they do during a full depolarization before they return to intracellular vestibule once the voltage drops. This behavior can be seen from Fig. 8(c), that *Q* reaches 1.57 at most which should be 2 instead if equilibrium, which can be reached if time is long enough, as shown in QV curve of Fig. 3(a). Fig. 8(d) shows that S4 initially farthest to the right lags behind R1 during movement and is almost caught up by R2. The maximum displacements of arginines and S4 calculated from Fig. 8(d) are $\Delta z_{1,CM}$ =1.36nm, $\Delta z_{2,CM}$ =0.966nm, $\Delta z_{3,CM}$ =0.459nm, $\Delta z_{4,CM}$ =0.316nm, and $\Delta Z_{4,CM}$ =0.616nm.

**5.6 Family of gating currents for a range of voltages**

Fig. 9(a) shows the time courses of gating current for a range of voltages *V*, ranging from -62mV to -8mV. We can see the area under gating current, for both on and off parts, increases with *V* since more arginines are transferred to extracellular vestibule as *V* increases. However, this area will saturate with further increasing *V* since all arginines would then be transferred to the extracellular vestibule. The shapes of this family of gating currents agree well with experiment [7] in both magnitude and time span. We can characterize the time span by fitting the time course (only the decay part) of a gating current by $ae^{-t/\tau_1} + be^{-t/\tau_2}$, $\tau_1 < \tau_2$, as generally conducted in experiment [7], where $\tau_1$ is the fast time constant and $\tau_2$ is the slow time constant. Usually the movement of arginines is dominated by $\tau_2$. Here $\tau_2$ was calculated from simulation and compared with experiment [7] as shown in Fig. 9(b). Since in our computation the time is in arbitrary unit, we have scaled the time to have the maximum $\tau_2$ to fit with its counterpart in experiment [7]. The trend of $\tau_2$ versus *V* in our result agrees well with experiment [7]. For the left branches to the maximum point in Fig. 9(b), simulation result fits very well with experiment. For the right branches to the maximum point, simulation result overestimates $\tau_2$ compared with experiment. This is consistent with the observation that the amount of transferred charges *Q* saturates faster in experiment data than in present simulation as *V* increases as shown in the QV curve of Fig. 3(a). This phenomenon is related to cooperativity of movement among arginines, that will be further discussed below.

**5.7 Effect of voltage pulse duration**

Fig. 10 shows the effect of voltage pulse duration with Fig. 10(a) for the case of partial depolarization and Fig. 10(b) for the case of full depolarization. Magnitude and time span of gating current are influenced by pulse duration in both cases, but the shape will asymptotically approach the same as pulse duration increases. This behavior occurs because driving arginines towards extracellular vestibule by the command potential takes time. If the pulse duration is long enough, the time course of *Q* will approach its steady state like in Fig. 8(a) for large depolarization. Partial depolarization takes longer time to reach its steady state as demonstrated in Fig. 8(c). Again these shapes of gating currents in Fig. 10 compare favorably with experiment [7].

**6. Conclusions**

The present 1D mechanical model of the voltage sensor tries to capture the essential structural details that are necessary to reproduce the basic features of experimentally recorded gating currents. After finding appropriate parameters, we find that the general kinetic and steady-state properties are well represented by the simulations. The continuum approach seems to be a good model of voltage sensors, provided that it takes into account all interactions, and satisfies conservation of current. The continuum approach used here describes the mechanical behavior of a single voltage sensor, but experimental measurements are ensemble averages of multiple ion channels (and hence multiple voltage sensors). There is no conflict, however. Experiments average of measurements of large numbers of voltage sensors as does our macroscopic model.

Here we simplify the profile of energy barrier in hydrophobic plug since the potential of mean force (PMF) in that region, and its variation with potential and conditions, is unknown. Therefore the next step is to model the details of interactions or the moving arginines with the wall of the hydrophobic plug. There is plenty of detailed information on the amino acid side chains in the plug and how each one of them has important effects in the kinetics and steady-state properties of gating charge movement [25]. The studied side chains reveal steric as well as dielectric interactions with the arginines that the present model does not have. On the other hand, the power of the present mathematical modeling is precisely the implementation of interactions, therefore we believe that when we add the dielectric details of the channel a better prediction of the currents should be attained. Also, here the plug energy barrier $V_b$ is assumed to be independent of time. However, once the first arginine enters the hydrophobic plug by carrying some water with it, this partial wetting of pore will lower $V_b$, chiefly consisting of solvation energy, and enable the next arginine to enter the plug with less difficulty. This might explain the cooperativity of movement among arginines when they jump through the plug. This remains to be included in the future modeling. Besides, it has been reported that a very strong electric field might affect the hydration equilibrium of the hydrophobic plug and would lower its hydration energy barrier as well [39]. This cooperativity of movement may help explain the quick saturation in the upper right branch of QV curve (and smaller $\tau_2$). This also remains to be considered in the future modeling.

Also, further work must address the mechanism of coupling between the voltage sensor movements and the conduction pore. It seems likely that the classical mechanical models will need to be extended to include coupling through the electrical field. It is possible that the voltage sensor modifies the stability of conduction current by triggering sudden transitions from closed to open state, in a controlled process reminiscent of Coulomb blockade.

**ACKNOWLEDGMENTS**

This research was sponsored in part by the Ministry of Science and Technology of Taiwan under grant no. MOST-102-2115-M-002-015-MY4 (T.L.H.) and MOST-104-2115-M-035 -002 -MY2 (T.L.H.). Dr. Horng also thanks the support of National Center for Theoretical Sciences Mathematical Division of Taiwan (NCTS/MATH), and Dr. Ren-Shiang Chen for the long-term benefiting discussions.

*Cell Dev. Biol.* 22:23-52.

[38] Trefethen, L. N. 2000. *Spectral Methods in MATLAB*. SIAM, Philadelphia.

[39] Vaitheeswaran, S., J. C. Rasaiah, and G. Hummer. 2004. Electric field and temperature effects on water in the narrow nonpolar pores of carbon nanotubes. *J. Chem. Phys.* 121(16):7955-7965.

[40] Vargas, E., F. Bezanilla, and B. Roux. 2011. In search of a consensus model of the resting state of a voltage-sensing domain. *Neuron*. 72(5):713-20.

[41] Varsos, K., J. Luntz, M. Welsh, and K. Sarabandi. 2011. Electric Field-Shaping Microdevices for Manipulation of Collections of Microscale Objects. *Proceedings of the IEEE*. 99(12): 2112-2124.

[42] Zhu, F., and G. Hummer. 2012. Drying transition in the hydrophobic gate of the GLIC channel blocks ion conduction. *Biophys. J.* 103:219-227.

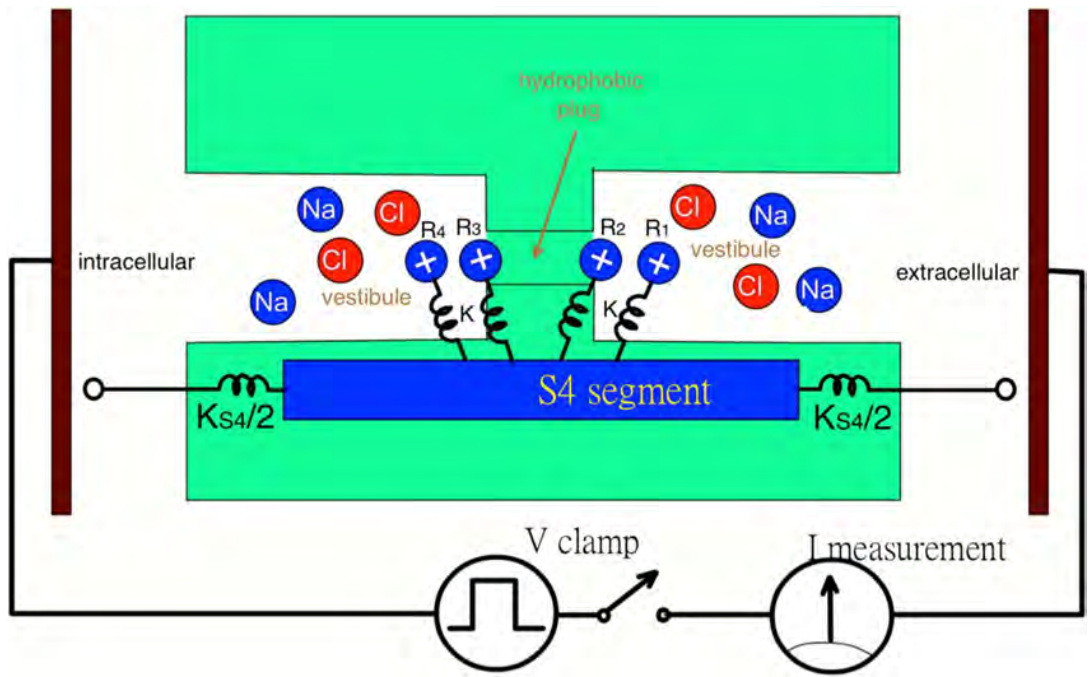

Figure 1. Geometric configuration of the model including the attachments of arginines to the S4 segment.

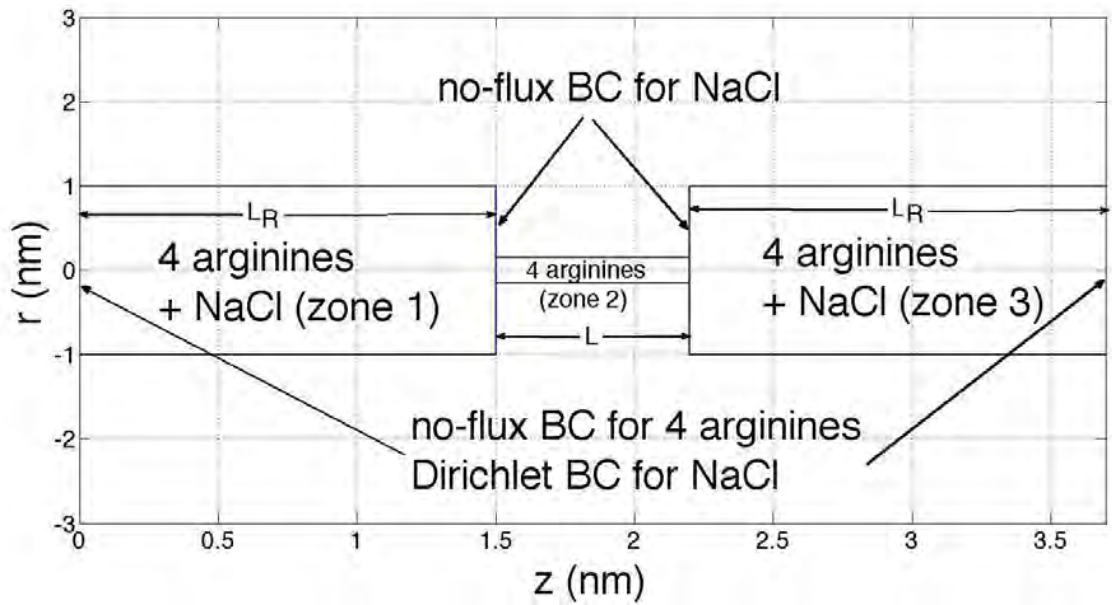

Figure 2. Following Figure 1, an axisymmetric 3-zone domain shape is designated in *r-z* coordinate for the current 1D model. Here the diameter of hydrophobic plug is 0.3nm (arginine's diameter); *L*=0.7nm; *L$_R$*=1.5nm; radius of vestibule is *R*=1nm. BC means Boundary Condition

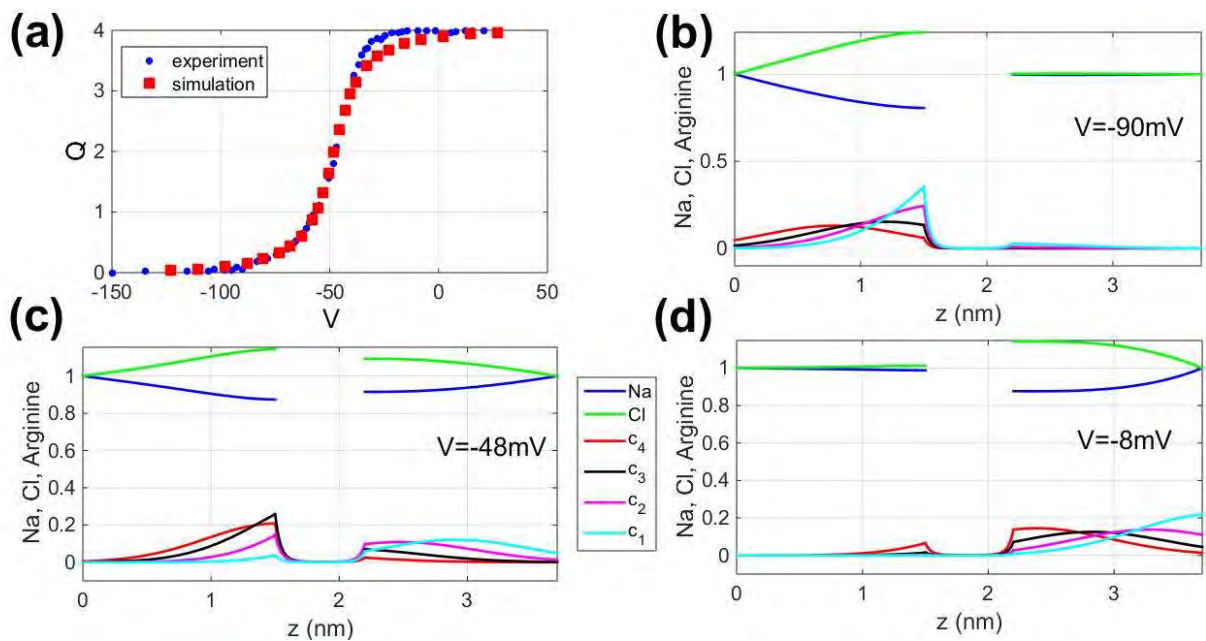

Figure 3. (a) QV curve and comparison with [7]. Steady-state distributions for Na, Cl and arginines at (b) *V*=-90mV, (c) *V*=-48mV, (d) *V*=-8mV.

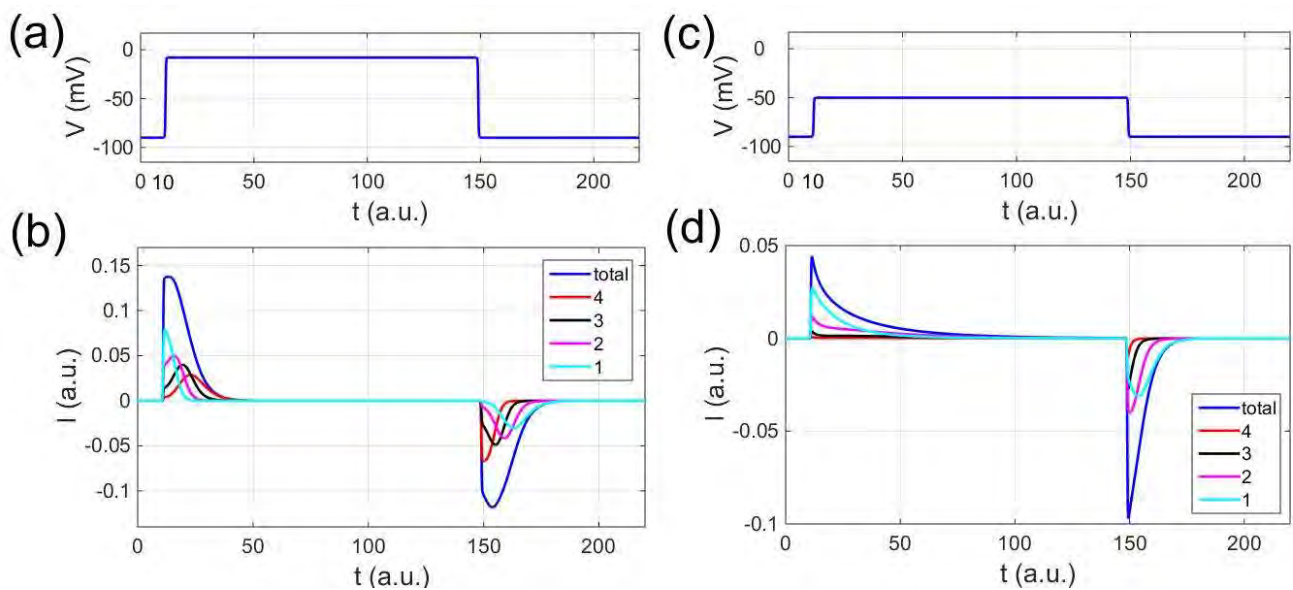

Figure 4. (a) Time course of $V$ rising from -90mV to -8mV at $t=10$, holds on till $t=150$, and drops back to -90mV, (b) time course of gating current and its components corresponding to (a), (c) time course of $V$ rising from -90mV to -50mV at $t=10$, holds on till $t=150$, and drops back to -90mV, (d) time course of gating current and its components corresponding to (c).

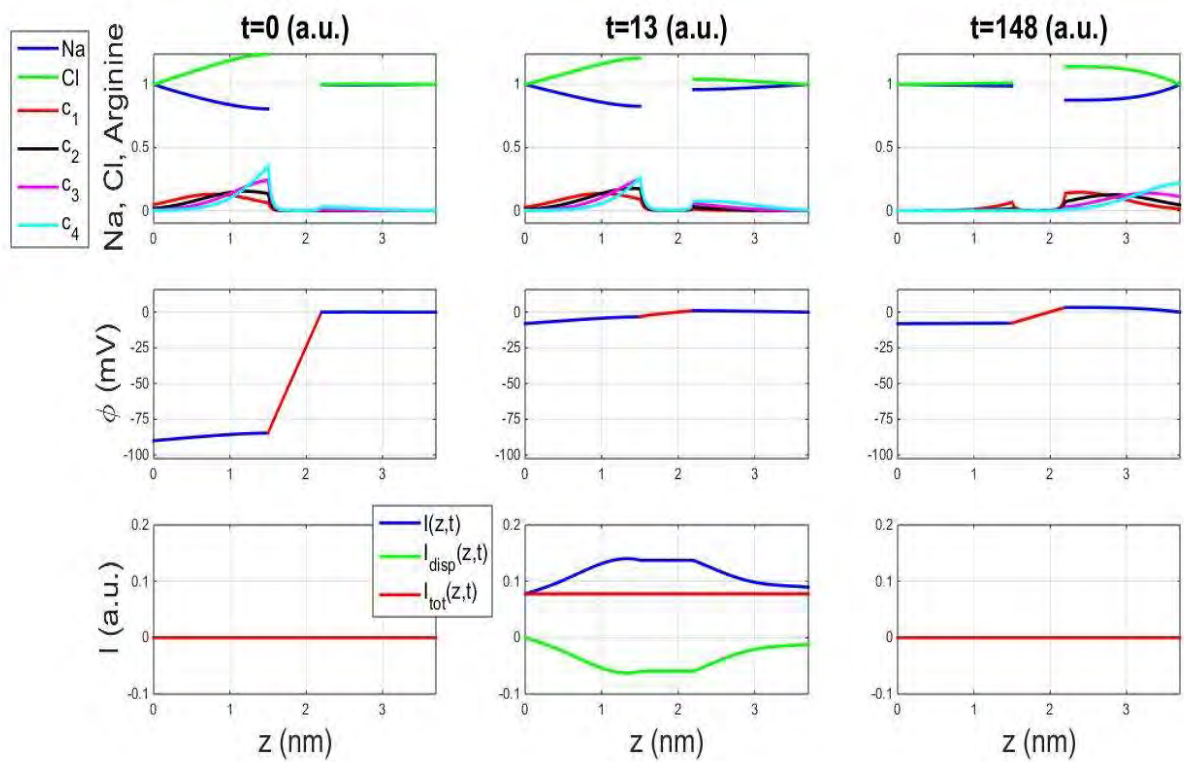

Figure 5. The four panels on the top row are species concentration distributions at $t=0, 13, 148$, for the case of large depolarization with $V$ rising from -90mV to -8mV at $t=10$, holding on till $t=150$, and dropping back to -90mV. The four panels on the middle row are concurrent electric potential profiles. The four panels on the bottom row are concurrent electric current profiles with components of flux of charge, displacement current and total current.

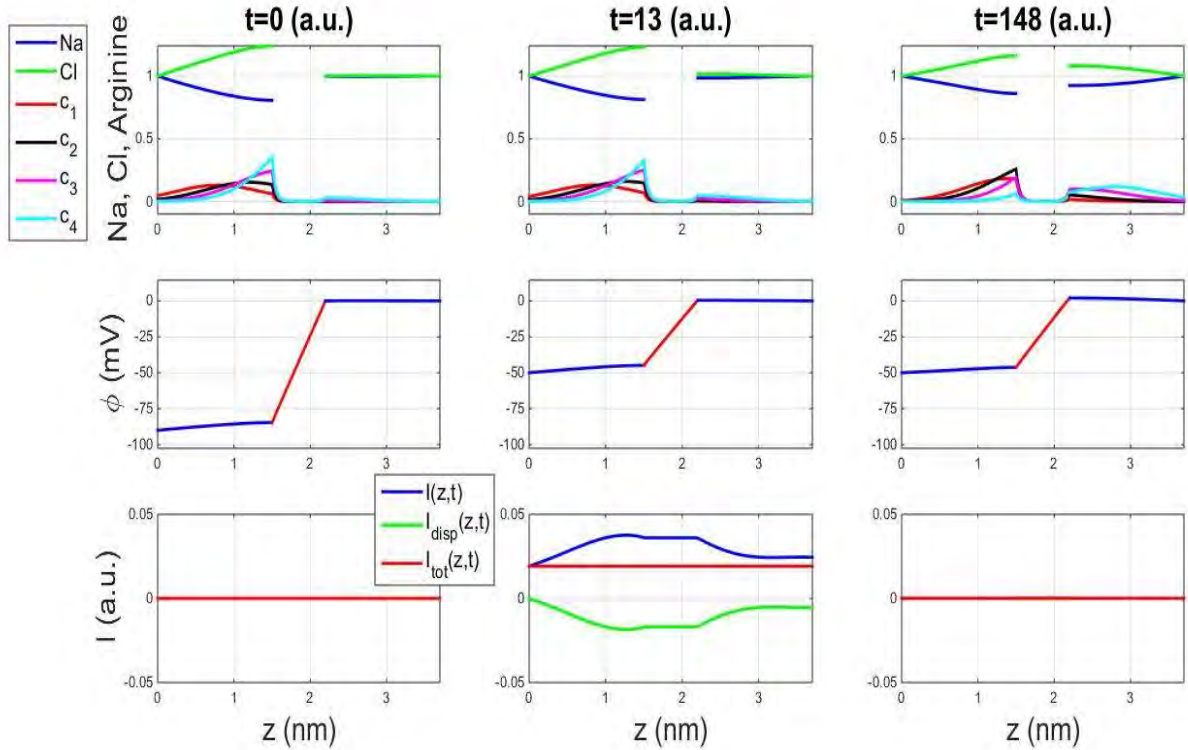

Figure 6. The four panels on the top row are species concentration distributions at $t$=0, 13, 148, for the case of large depolarization with $V$ rising from -90mV to -50mV at $t$=10, holding on till $t$=150, and dropping back to -90mV. The four panels on the middle row are concurrent electric potential profiles. The four panels on the bottom row are concurrent electric current profiles with components of flux of charge, displacement current and total current.

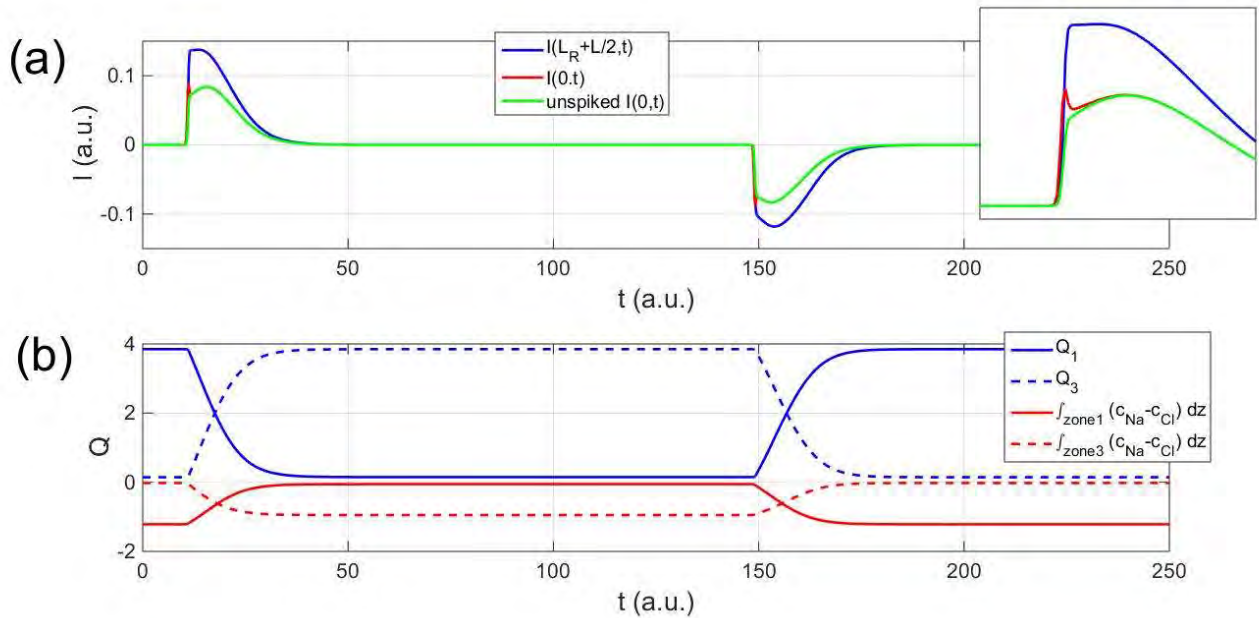

Figure 7. (a) Time courses of $I\left(L_R + \frac{L}{2}, t\right)$, $I(0,t)$ and unspiked $I(0,t)$ for the case of full depolarization with $V$ rising from -90mV to -8mV at $t$=10, holding on till $t$=150, and dropping back to -90mV. The inset plot is a blow-up of ON-current to visualize the difference of $I(0,t)$ and unspiked

$I(0, t)$ more clearly. (b) Time courses of $Q_1, Q_3, \int_0^{L_R}(c_{Na} - c_{Cl})\,dz$, and $\int_{L_R+L}^{2L_R+L}(c_{Na} - c_{Cl})\,dz$ under the same depolarization scenario as (a).

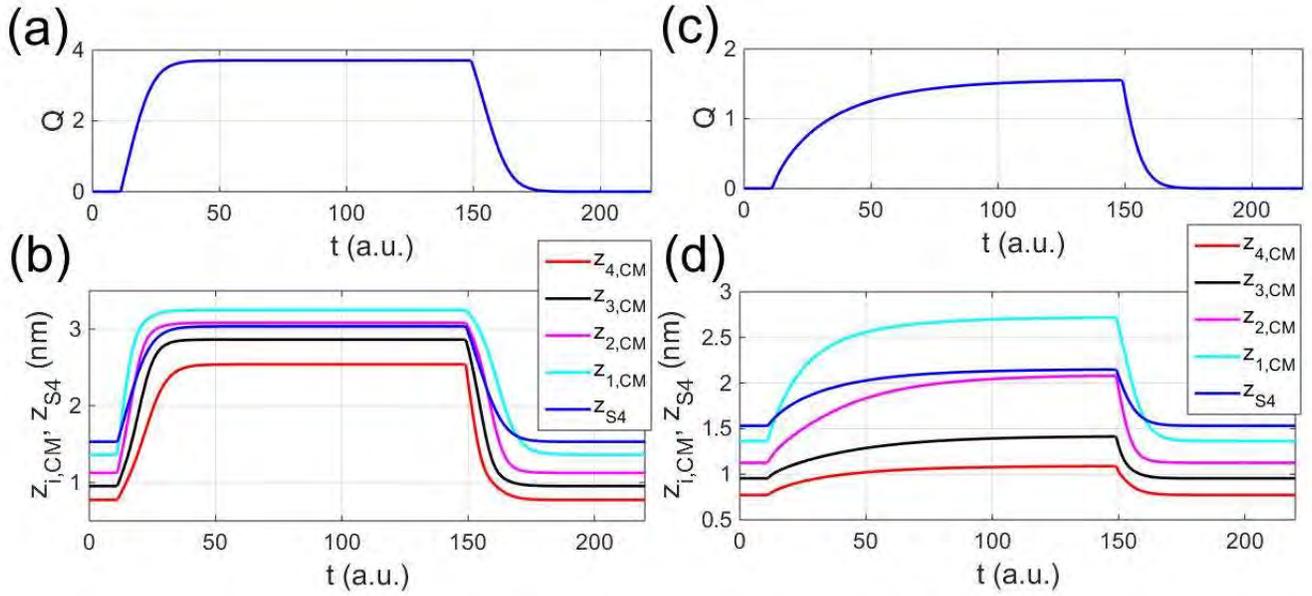

Figure 8. (a) and (c) are time courses of amount of arginines moved to extracellular vestibule. (b) and (d) are center-of-mass trajectories of individual arginines and S4. (a) and (b) are the case of large depolarization with $V$ rising from -90mV to -8mV at $t$=10, holding on till $t$=150, and dropping back to -90mV. (c) and (d) are the case of partial depolarization with $V$ rising from -90mV to -50mV at $t$=10, holding on till $t$=150, and dropping back to -90mV.

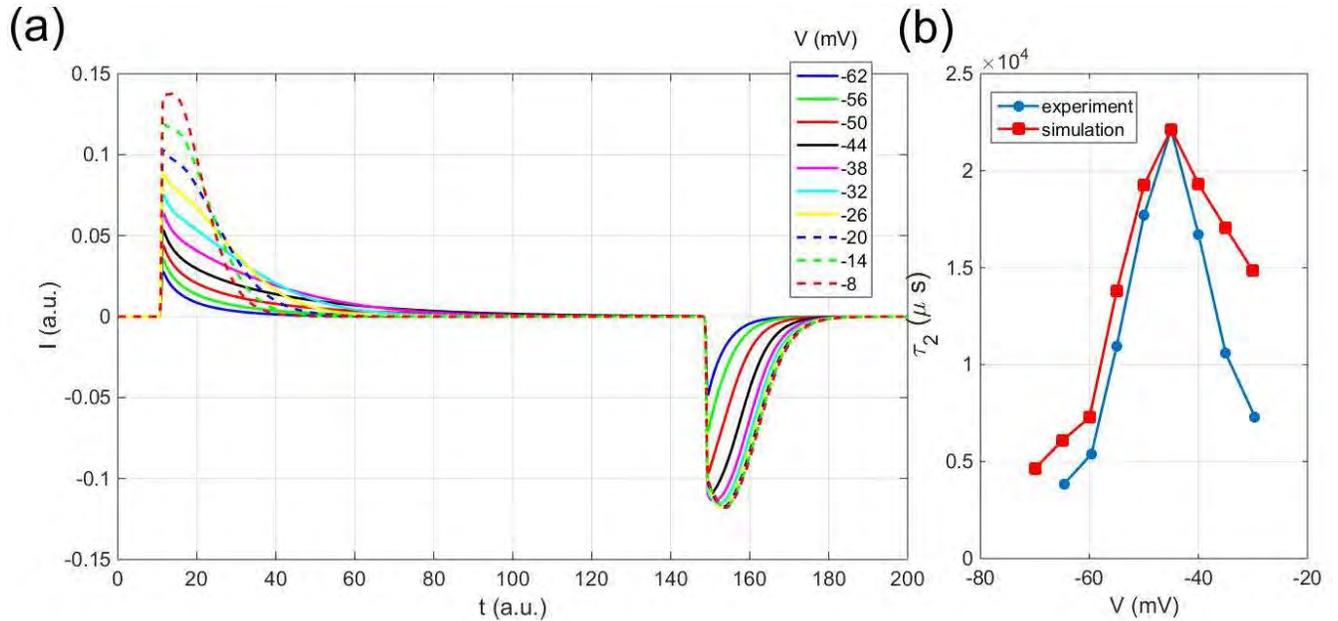

Figure 9. (a) Time courses of gating current with voltage rising from -90mV to $V$ mV at $t$=10, holds on till $t$=150, and drops back to -90mV, where $V$=-62, -50, … -8 mV. (b) $\tau_2$ versus $V$ compared with experiment [7].

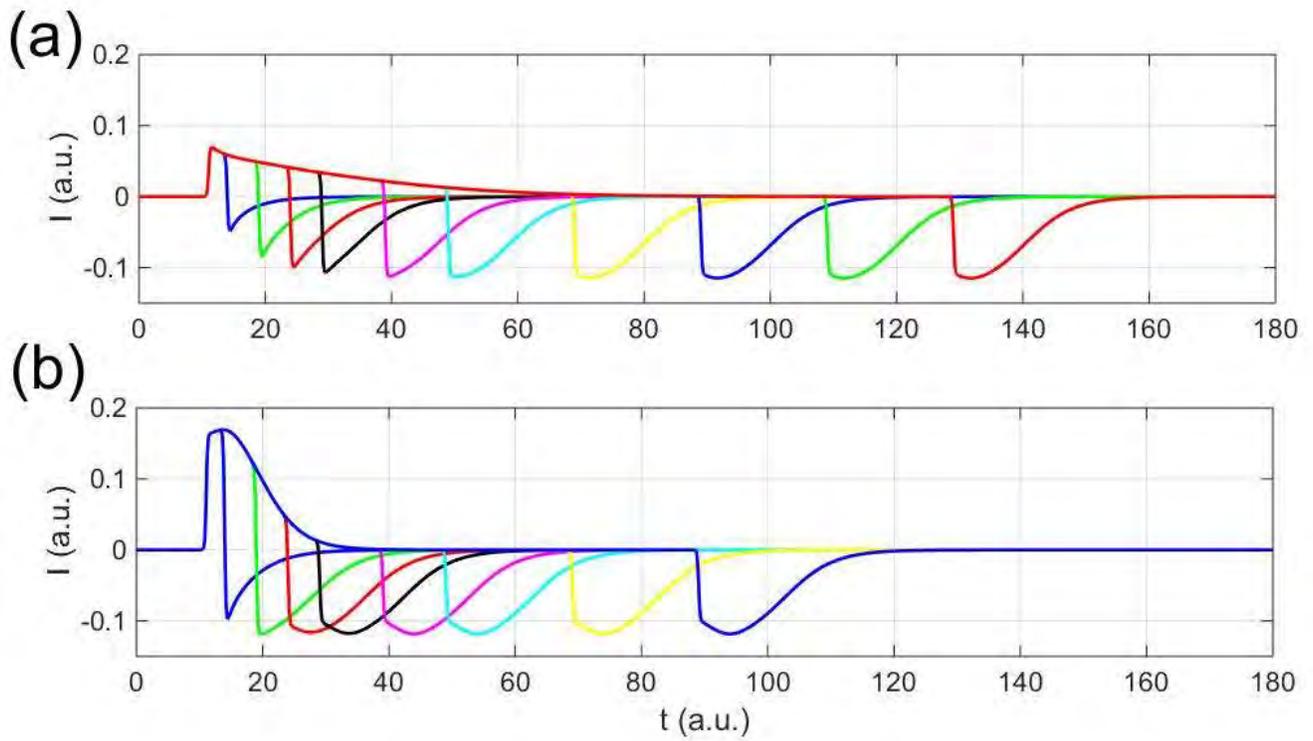

Figure 10. Effect of voltage pulse duration: (a) *V* increases from -90mV to -35mV at *t*=10 and drops back to -90mV at various times, (b) *V* increases from -90mV to 0mV at *t*=10 and drops back to -90mV at various times.